\documentclass[11pt,twoside]{article}

\usepackage{asp2006}
\usepackage{lscape}
\usepackage{graphicx}

\begin{document}
\title{Supergranulation Scale Connection Simulations}
\author{Robert F. Stein\altaffilmark{1}, {\AA}ke Nordlund\altaffilmark{2}, Dali Georgoviani\altaffilmark{1}, David Benson\altaffilmark{3}, Werner Schaffenberger\altaffilmark{1}}   
\altaffiltext{1}{Michigan State University, East Lansing, MI 48824, USA}    
\altaffiltext{2}{Niels Bohr Institute, Copenhage, DK-2500, DK}
\altaffiltext{3}{Kettering University, Flint, MI}

\begin{abstract} 
Results of realistic simulations of solar surface convection on the
scale of supergranules (96 Mm wide by 20 Mm deep) are presented.
The simulations cover only 10\% of the geometric depth of the solar
convection zone, but half its pressure scale heights.  They include
the hydrogen, first and most of the second helium ionization zones.
The horizontal velocity spectrum is a power law and the horizontal
size of the dominant convective cells increases with increasing
depth.  Convection is driven by buoyancy work which is largest close
to the surface, but significant over the entire domain.  Close to
the surface buoyancy driving is balanced by the divergence of the
kinetic energy flux, but deeper down it is balanced by dissipation.
The damping length of the turbulent kinetic energy is 4 pressure
scale heights.  The mass mixing length is 1.8 scale heights.  Two
thirds of the area is upflowing fluid except very close to the
surface.  The internal (ionization) energy flux is the largest
contributor to the convective flux for temperatures less than 40,000
K and the thermal energy flux is the largest contributor at higher
temperatures.   This data set is useful for validating local
helioseismic inversion methods.  Sixteen hours of data are available
as four hour averages, with two hour cadence, at
steinr.msu.edu/$\sim$bob/96averages, as idl save files.  The variables
stored are the density, temperature, sound speed, and three velocity
components.  In addition, the three velocity components at 200 km
above mean continuum optical depth unity are available at 30 sec.
cadence.
\end{abstract}

\section{The Simulation}
Solar surface convection on supergranular scales (96 Mm wide by 20
Mm deep) was simulated by solving the conservation equations for
mass, momentum and internal energy.  Spatial derivatives were
evaluated using sixth order finite differences \citep{Nagarajan2003}
and the time advance was by a low memory, third order Runge-Kutta
scheme \citep{Kennedy1999}.  The calculations were performed on a
grid of $1000^2 \times 500$ giving a resolution of 96 km horizontally
and 12-70 km vertically.  f-plane rotation is included corresponding
to a latitude of 30.

A tabular equation of state which includes local thermodynamic
equilibrium (LTE) ionization of the abundant elements as well as
hydrogen molecule formation, was used to obtain the pressure and
temperature as a function of log density and internal energy per
unit mass.  The radiative heating/cooling was obtained by solving
the radiation transfer equation in both continua and lines using
the Feautrier method, assuming Local Thermodynamic Equilibrium
(LTE).  The number of wavelengths for which the transfer equation
is solved is drastically reduced by using a multi-group method
whereby the opacity at each wavelength is placed into one of four
bins according to its magnitude and the source function is binned
the same way \citep{Nordlund1982,Stein2003}.

Horizontal boundary conditions are periodic, while top and bottom
boundary conditions are open.  The code is stabilized by diffusion in
the momentum and energy equations, using a variable diffusivity
that depends on the sound speed, the fluid velocity and the
compression.

\section{The Atmosphere}

\begin{figure}
  \centerline{
  \includegraphics[width=.5\textwidth]{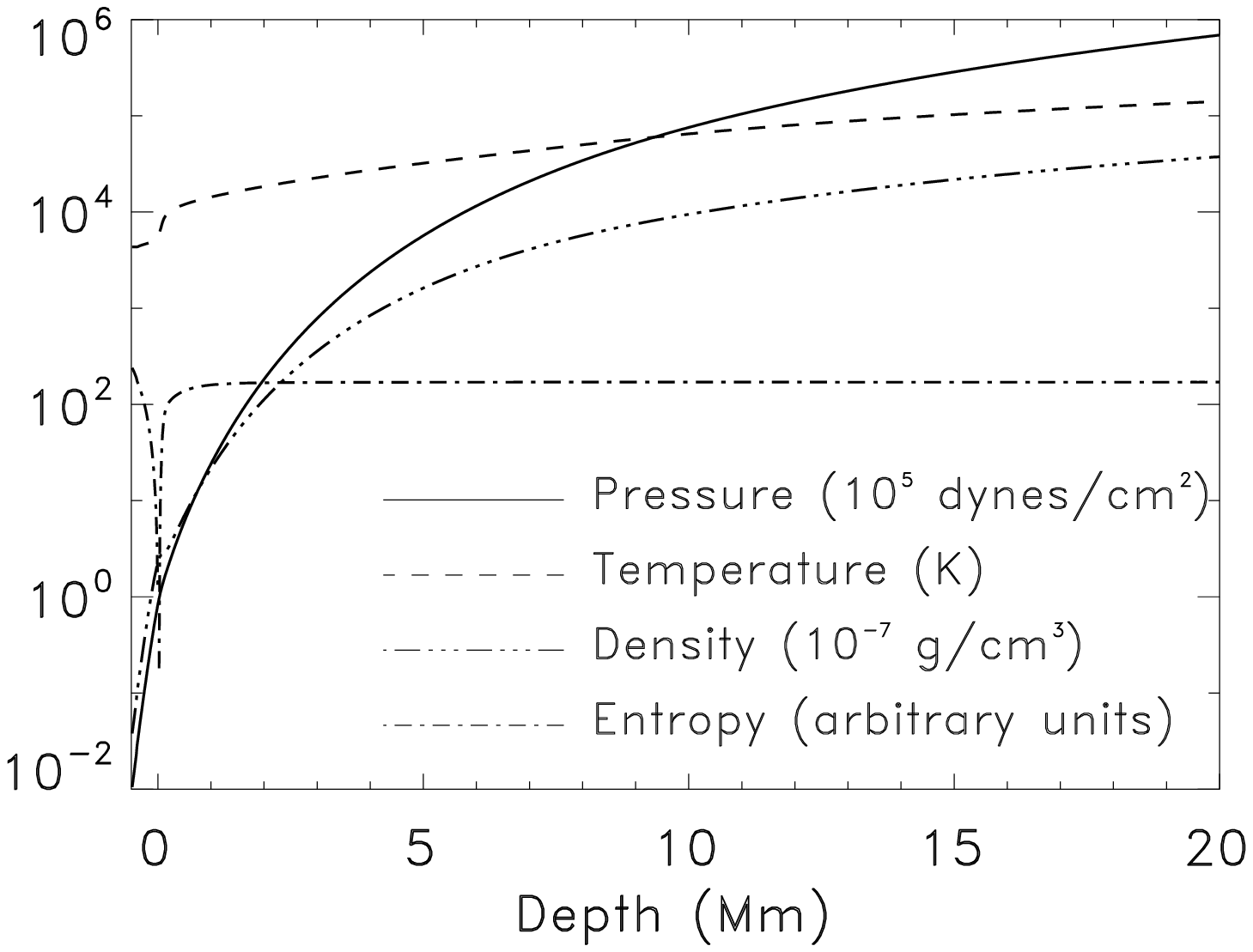}
  \includegraphics[width=.5\textwidth]{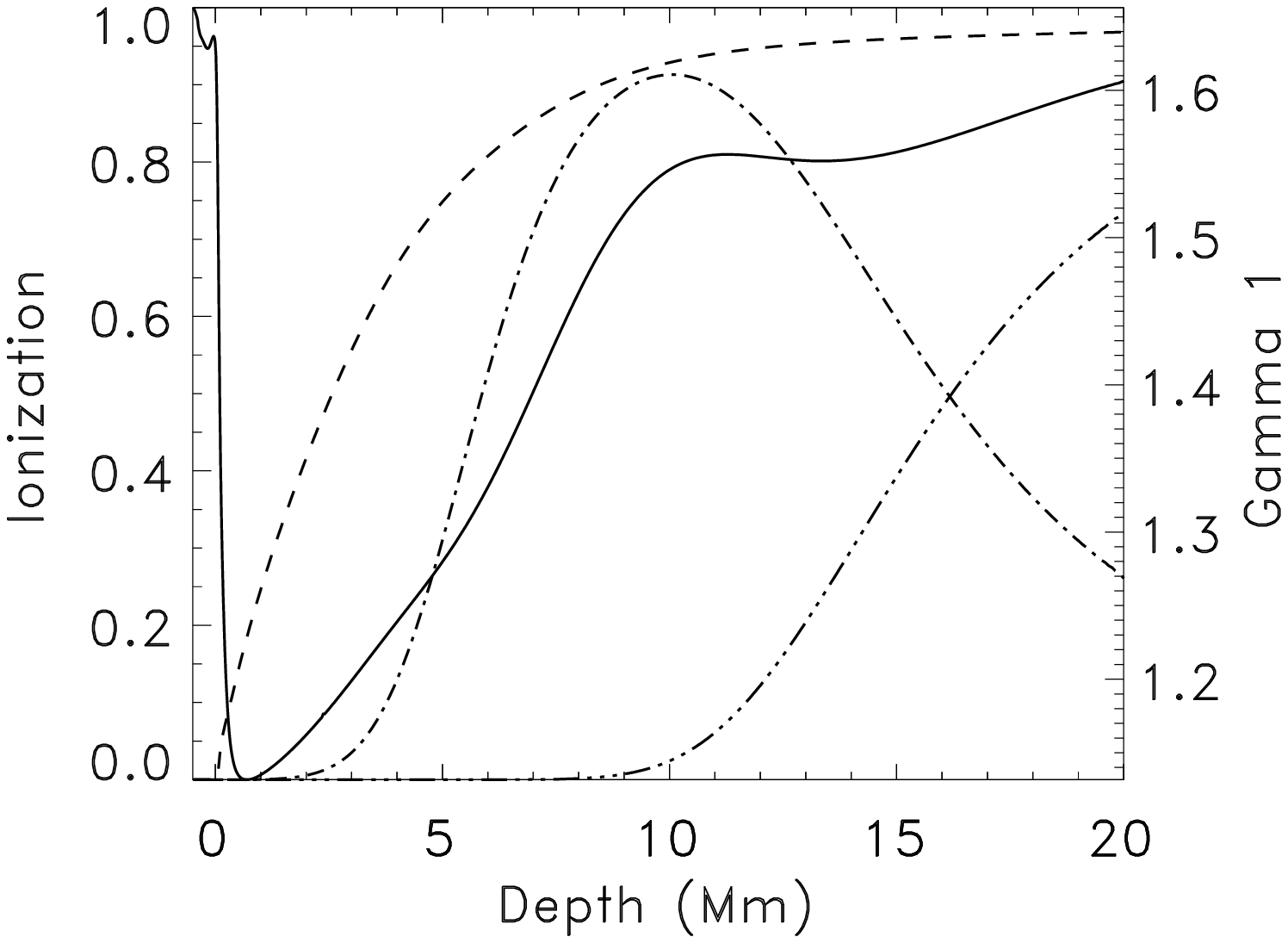}
  }
  \caption{Pressure, temperature, density, entropy (left), Gamma1
  (solid), H ionization fraction (dashed), and He ionization fractions
  (dash dot and dast dot dot dot) (right).  The computational domain
  covers 10 \% of the geometric depth of the solar convection zone and
  half its scale heights.  There are 5 orders of magnitude in pressure,
  four in density and one in temperature within the convective region
  included.}
  \label{fig:atmosphere}
\end{figure}

The computational domain covers five orders of magnitude in pressure in
the convective region, which is half the scale heights in the entire
solar convection zone, even though the domain is only 10\% of the zones
geometric depth (Fig. \ref{fig:atmosphere}).  This includes the
hydrogen ionization layer and the first and most of the second helium
ionization layers.  A shear layer develops with a shear of 100 m/s
across the 20 Mm depth of the domain.  

\begin{figure}
   \centerline{\includegraphics[width=0.5\textwidth]{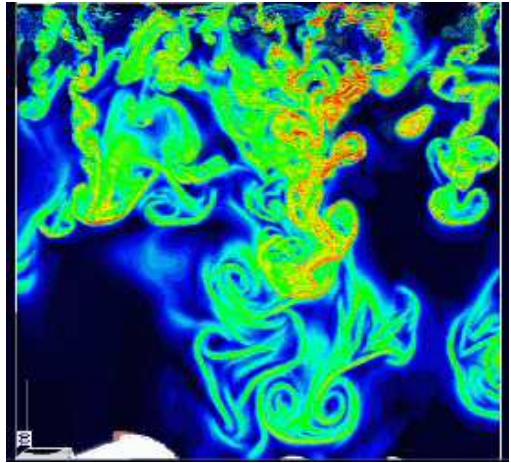}}
   \caption{Finite time (11.75 hours) Lyapunov exponent field ov a
   subdomain 21 Mm wide by 19 Mm high, by 0.5 Mm thick, from a 48
   wide $\times$ 20 Mm deep simulation.  This corresponds closely
   to the magnitude of the vorticity. Figure courtesy Bryan Green,
   AMTI/NASA.}
   \label{fig:vorticity}
\end{figure}

\begin{figure}
   \centerline{\includegraphics[width=0.8\textwidth,angle=90]{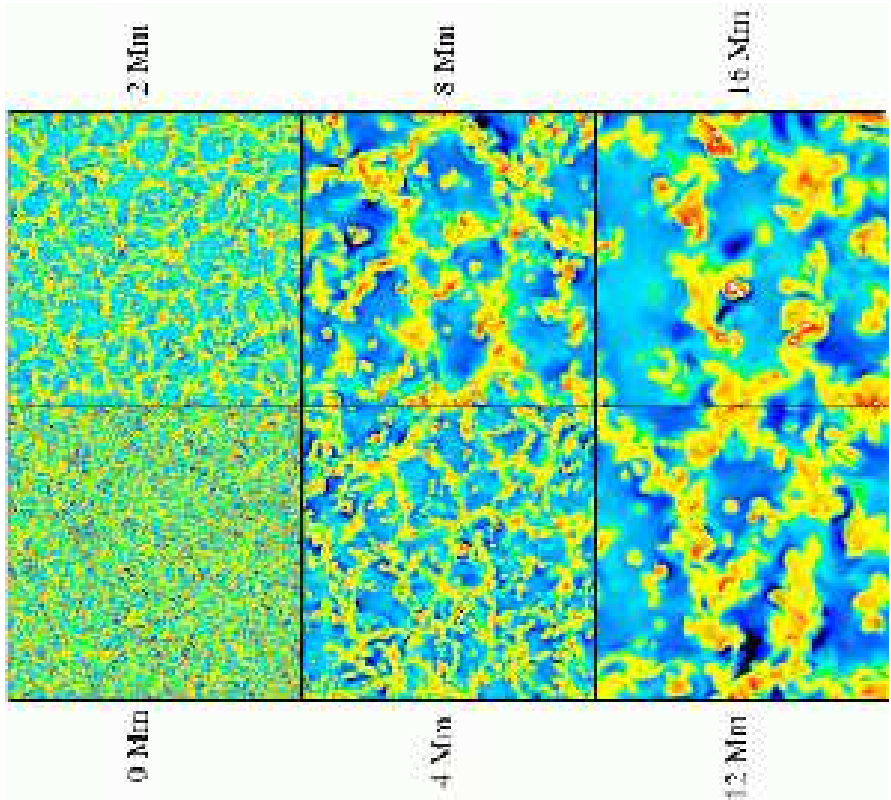}}
   \caption{Vertical velocity on horizontal slices at the surface
   and 2, 4, 8, 12, and 16 Mm below the surface.  Red and yellow are
   downflows.  Blue and green are upflows.  The dominant horizontal 
   scale of the convection increases monotonically with increasing 
   depth.}
   \label{fig:scale}
\end{figure}

Convective motions are turbulent.  Descending vortex rings have
typical mushroom shape and are connected back to surface with
multiple twisted vortex tubes (Fig.~\ref{fig:vorticity}).  The
horizontal scale of the convection increases from granule size at
the surface to supergranule size near the bottom of the computational
domain (Fig.~\ref{fig:scale}).  The horizontal velocity spectrum
at the surface is a power law, with the magnitude of the velocity
decreasing linearly with increasing size.  There is a peak at granule
scales and then a fall off at still smaller scales.  About $2/3$
of the area is upflows and $1/3$ downflows nearly independent of
depth.

\smallskip

\section{Driving and Damping}

\begin{figure}
  \centerline{
  \includegraphics[width=.5\textwidth]{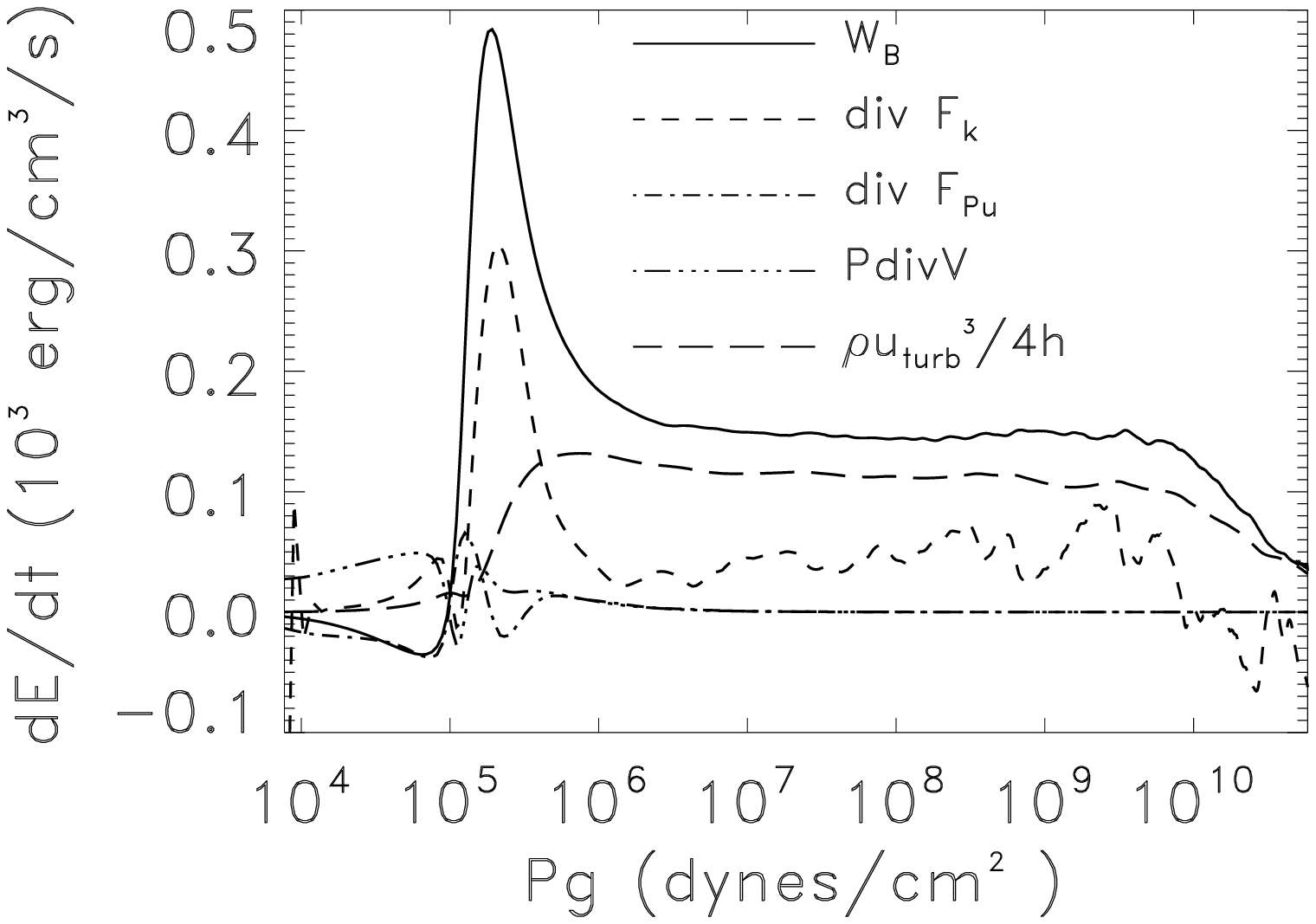}
  \includegraphics[width=.5\textwidth]{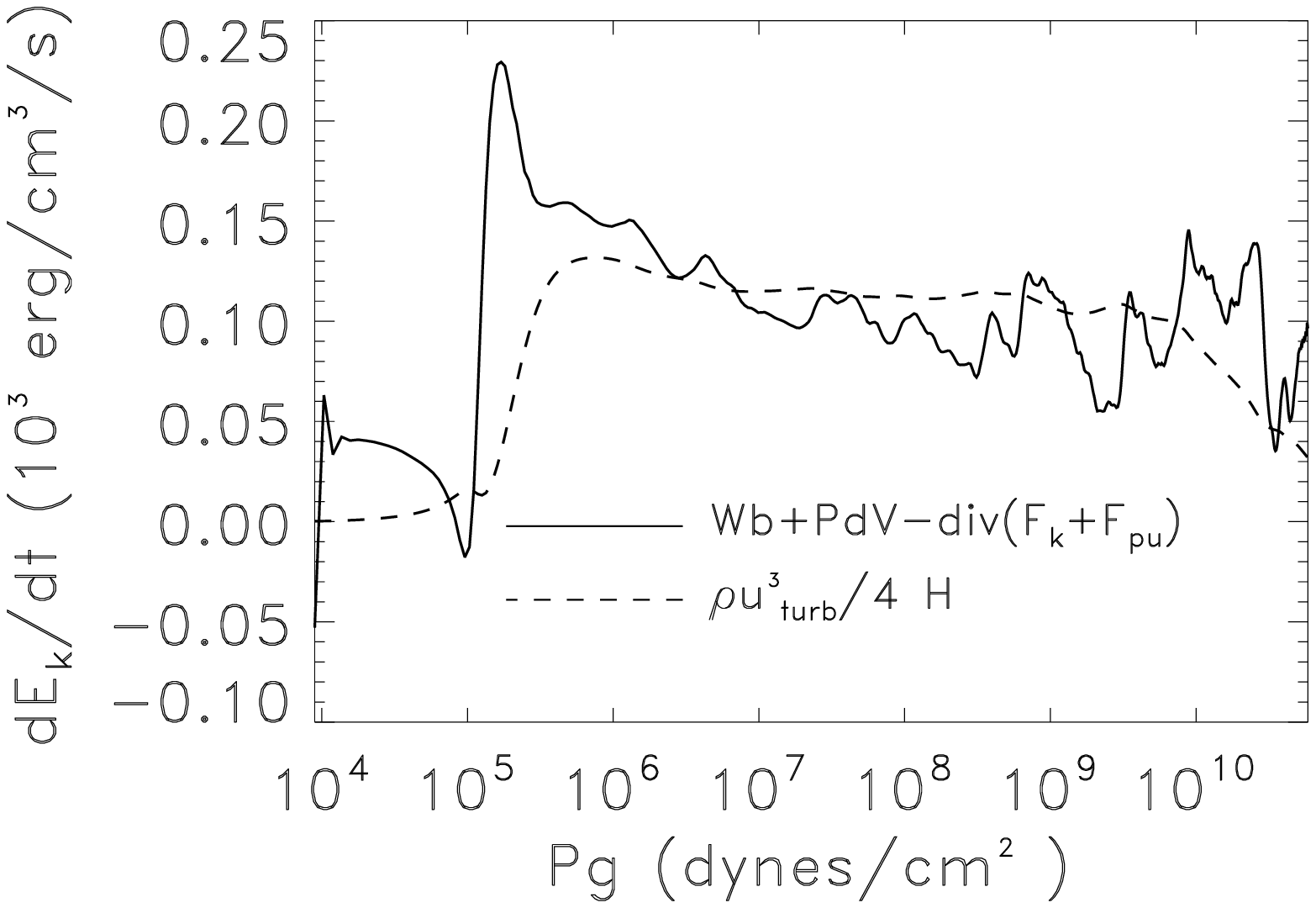}
  }
  \caption{Contributions to the kinetic energy equation.  Most of the
  buoyancy driving occurs just below the surface where the entropy
  fluctuations are large, but remains significant and nearly constant at
  larger depths.  The buoyant work is balanced by the divergence of the
  kinetic energy flux near the surface and by dissipation in the
  interior.  The change in behavior near the bottom boundary is due to
  boundary condition effects and is not physical.}
  \label{fig:drivedamp}
\end{figure}

Convection on the Sun is driven primarily by radiative cooling in
a thin boundary layer at the bottom of the photosphere where radiation
begins to escape to space.  Radiative cooling produces low entropy,
over dense fluid that is pulled down by gravity.  There is also
driving from the bottom of the convection zone where absorption of
the radiative flux heats the fluid producing high entropy, under
dense fluid.  However, the entropy fluctuations near the surface
are much much larger than those near the bottom of the convection
zone, so most of the buoyant work occurs near the surface
(Fig.~\ref{fig:drivedamp}).  The buoyancy driving is balanced by
the divergence of the kinetic energy flux close to the surface and
by dissipation at small scales through the remainder of the convection
zone.  The damping length is 4 pressure scale heights (Fig.
\ref{fig:drivedamp}).

\section{Energy Fluxes}

\begin{figure}
  \centerline{
  \includegraphics[width=.5\textwidth]{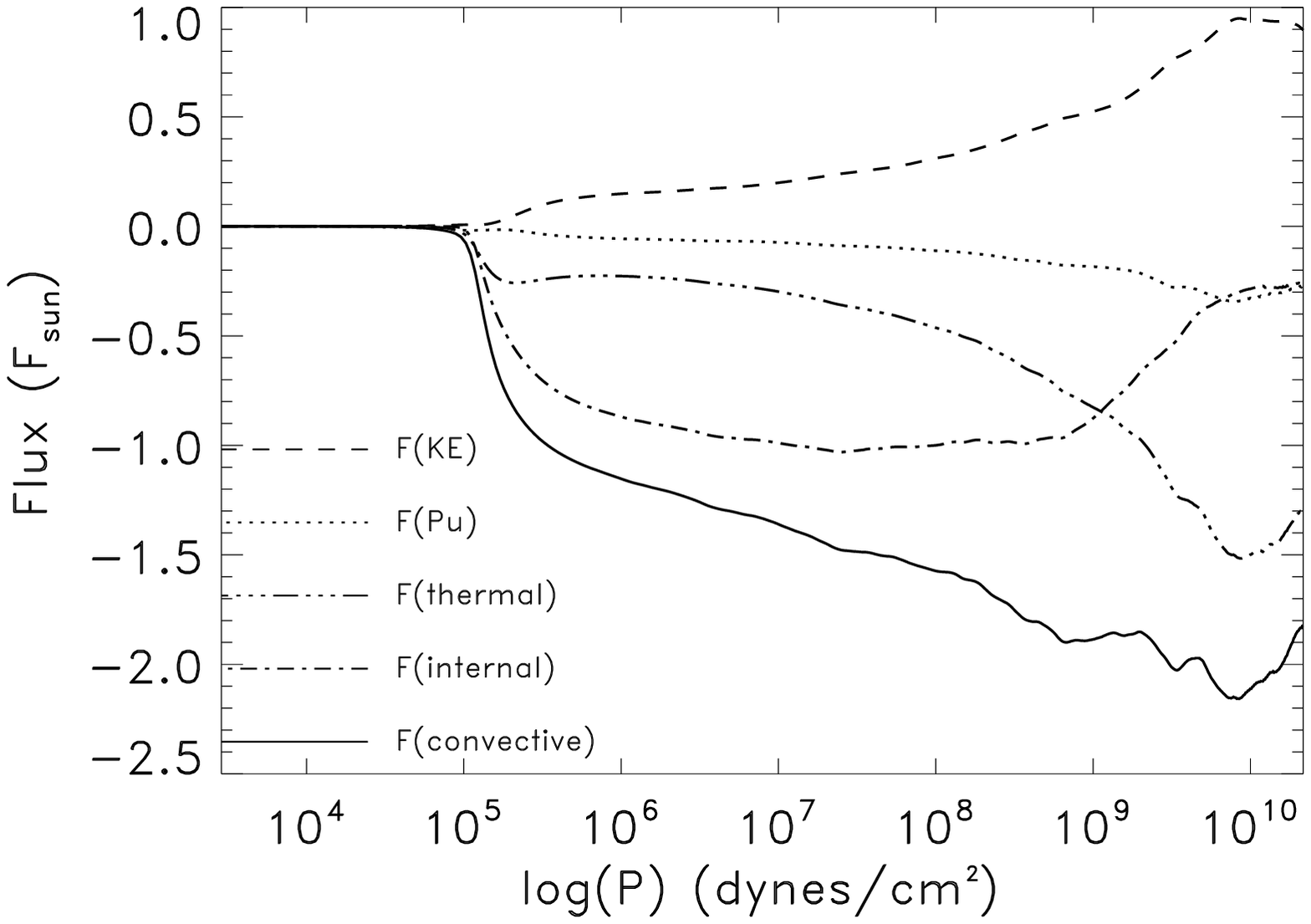}
  \includegraphics[width=.5\textwidth]{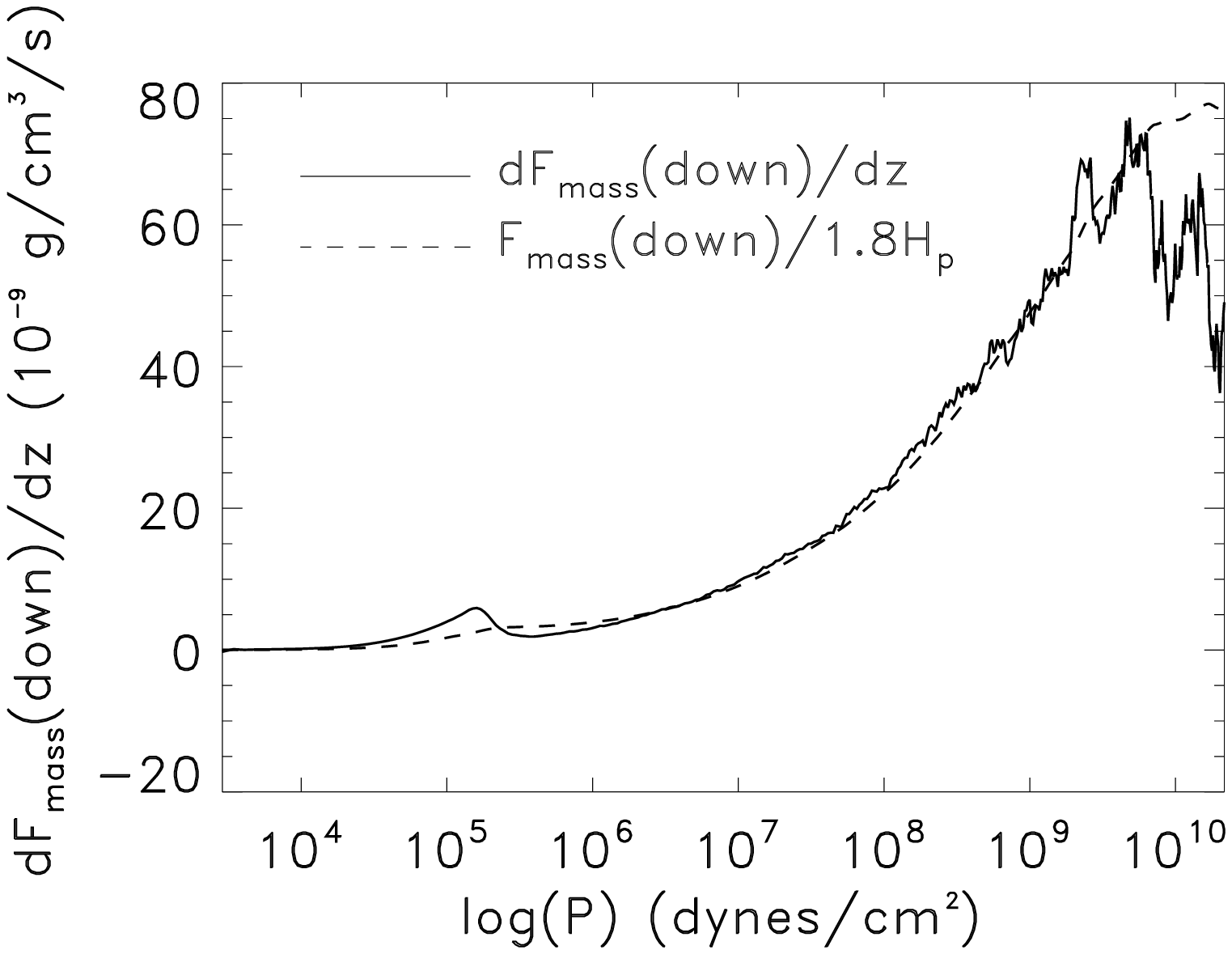}
  }
  \caption{Energy fluxes (left) and mass mixing length (right).  Note
  negative fluxes are upward energy transport.  The kinetic energy flux
  is downward and approximately half the magnitude of the total other,
  upward, energy fluxes. Energy is transported primarily as the
  internal (ionization) energy below 40,000 K and primarily as the
  thermal energy at higher temperatures.  Upflowing fluid turns over
  and is entrained in the downflows in order to conserve mass with a
  length scale or mixing length of 1.8 pressure scale heights.
  }

  \label{fig:fluxmix}
\end{figure}

Most of the energy in the cooler part of the convection zone close
to the surface where hydrogen is partially ionized is carried by
the internal (ionization) energy flux.  At temperatures above 40,000
K, this decreases and most of the energy is transported by the
thermal energy flux (Fig. \ref{fig:fluxmix}).

\section{Mixing Length}

The mass mixing length is the inverse of the logarithmic
derivative of the unidirectional (either up or down) mass flux,
$\ell = \left(d F_{\rm mass}(down)/dr / F_{\rm mass}(down) \right)^{-1}$.
Figure \ref{fig:fluxmix} shows that over most of the computational domain
except right near the surface the mass mixing length is 1.8 pressure
scale heights.  This is an easy, unambiguous, calculation to make
for other stars as well to determine the appropriate mixing length.

\section{Application to Local Helioseismology}

Sixteen hours of data are available as four hour averages, with two
hour cadence, at steinr.msu.edu/$\sim$bob/96averages, as idl save files.
The variables stored are the density, temperature, sound speed,
three velocity components.  In addition, the three velocity components
at 200 km above mean continuum optical depth unity are available
at 30 sec. cadence.

\section{Summary}

A realistic simulation of supergranulation scale convection, covering
half the scale heights in the solar convection zone can be used to
evaluate local helioseismic inversion proceedures.  It has also
allowed us to determine the mass mixing length (1.8 $H_P$), to show
that most of the energy is transported as ionization energy below
40,000 K and as thermal energy above that temperature, and that the
buoyancy driving is largest close to the surface, but significant
throughout the simulated domain.  In the interior, this driving is
balanced primarily by dissipation at small scales and the damping
length is 4 $H_P$.  Four hour averages with two hour candence of
the density, temperature, sound speed, and three velocity components
are available as idl save files.  In addition, the three velocity
components at 200 km above mean continuum optical depth unity are
available at 30 sec. cadence.

\acknowledgements 
The calculations were performed at the NASA High End Computing
Columbia supercomputer.  Support for this project was provided by
NASA grants NNX07AO71G, NNX07AH79G, NNX07AI08G and NNX08AH44G, and
NSF grant AST 0605738.  This support is greatly appreciated.


\begin{thebibliography}{}

\bibitem[Kennedy, Carpenter \& Lewis (1999)]{Kennedy1999}
C.~A. Kennedy, M.~H. Carpenter, \& R.~M. Lewis 1999, 
ICASE Report No. 99-22, NASA/CR-1999-209349

\bibitem[Nagarajan, Lele \& Ferziger (2003)]{Nagarajan2003}
S. Nagarajan, S.~K. Lele \& J.~H. Ferziger 2003, {J. Comp. Phys.},
{191}, 392--429

\bibitem[Nordlund (1982)]{Nordlund1982}
{\AA}. Nordlund, {Astron. \& Astrophys.} 1982, {107}, 1--10

\bibitem[Stein \& Nordlund (2003)]{Stein2003}
R.~F. Stein and {\AA}. Nordlund 2003, in {Stellar Atmosphere Modeling}, eds. I. Hubeny, D. Mihalas and K. Werner, ASP Conf. Proc. 288, 519--532

\end{thebibliography}
\end{document}